\documentclass[twocolumn,epjc3]{svjour3}

\usepackage{amsmath}
\usepackage{euscript,amssymb,epsfig}
\usepackage{graphicx}
\usepackage{amsfonts,latexsym}

\RequirePackage{graphicx}

\newcommand{\be}[1]{\begin{equation}\label{#1}}
\newcommand{\ee}{\end{equation}}
\newcommand{\ba}[1]{\begin{eqnarray}\label{#1}}
\newcommand{\ea}{\end{eqnarray}}
\newcommand{\rf}[1]{(\ref{#1})}
\newcommand{\nn}{\nonumber}

\journalname{Eur. Phys. J. C}

\begin{document}

\title{Lattice Universe: examples and problems}

\author{Maxim Brilenkov\thanksref{e1,addr1} \and Maxim Eingorn\thanksref{e2,addr2} \and Alexander Zhuk\thanksref{e3,addr3}}

\thankstext{e1}{e-mail: maxim.brilenkov@gmail.com}

\thankstext{e2}{e-mail: maxim.eingorn@gmail.com}

\thankstext{e3}{e-mail: ai.zhuk2@gmail.com}

\institute{Department of Theoretical Physics, Odessa National University,\\ Dvoryanskaya st. 2, Odessa 65082, Ukraine\\ \label{addr1} \and Physics Department,
North Carolina Central University,\\ Fayetteville st. 1801, Durham, North Carolina 27707, U.S.A.\\ \label{addr2} \and Astronomical Observatory, Odessa National
University,\\ Dvoryanskaya st. 2, Odessa 65082, Ukraine \label{addr3}}

\date{Received: date / Accepted: date}

\maketitle

\begin{abstract}
We consider lattice Universes with spatial topologies $T\times T\times T$, $\; T\times T\times R\; $ and $\; T\times R\times R$. In the Newtonian limit of
General Relativity, we solve the Poisson equation for the gravitational potential in the enumerated models. In the case of point-like massive sources in the
$T\times T\times T$ model, we demonstrate that the gravitational potential has no definite values on the straight lines joining identical masses in neighboring
cells, i.e. at points where masses are absent. Clearly, this is a nonphysical result since the dynamics of cosmic bodies is not determined in such a case. The
only way to avoid this problem and get a regular solution at any point of the cell is the smearing of these masses over some region. Therefore, the smearing of
gravitating bodies in $N$-body simulations is not only a technical method but also a physically substantiated procedure. In the cases of $\; T\times T\times
R\; $ and $\; T\times R\times R$ topologies, there is no way to get any physically reasonable and nontrivial solution. The only solutions we can get here are
the ones which reduce these topologies to the $T\times T\times T$ one.
\end{abstract}

\maketitle

\vspace{.5cm}


\section{Introduction}

\setcounter{equation}{0}

Papers devoted to the lattice Universe can be divided into two groups. The first group includes articles (see, e.g.,
\cite{LinWh,KNic,CF1,CF2,Clift,UELar,CRT,YANT,FlorLeon,BruLar1,BruLar2}) offering alternative cosmological models. Despite the great success of the standard
$\Lambda$CDM model, it has some problematic aspects. The main one is the presence of dark energy and dark matter which constitute about 96\% of the total
energy density in the Universe. However, the nature of these components is still unknown. Another subtle point is that the conventional model is based on the
Friedmann-Lemaitre-Robertson-Walker (FLRW) geometry with the homogeneous and isotropic distribution of matter in the form of a perfect fluid. Observations show
that such approximation works well on rather large scales. According to simple estimates made on the basis of statistical physics, these scales correspond to
190 Mpc \cite{EZcosm2} which is in good agreement with observations. This is the cell of uniformity size. Deep inside this cell, our Universe is highly
inhomogeneous. Here, we clearly see galaxies, dwarf galaxies, groups and clusters of galaxies. Therefore, it makes sense to consider matter on such scales in
the form of discrete gravitational sources. In this case, we arrive at the question how this discrete distribution influences global properties and dynamics of
the Universe. This problem was investigated in the above mentioned papers (see also \cite{CELU}). Here, gravitating masses are usually distributed in a very
simplified and artificial way. They form either periodic structures of identical masses with proper boundary conditions or correspond to Einstein equation
solutions (e.g., Schwarzschild or Schwarzschild-de Sitter solutions) matching with each other with the help of the Israel boundary conditions. Usually, such
models do not rely on the $\Lambda$CDM background solution and do not include observable parameters (e.g., the average rest-mass density $\bar{\rho}$ of matter
in the Universe). As a result, these models have nothing common with the observable Universe. Their main task is to find new phenomena following from
discretization and nontrivial topology.

Papers from the second class are devoted to numerical $N$-body simulations of the observable Universe.  Here, the lattice is constructed as follows. In the
spatially flat Universe, we choose a three-dimensional cell with $N$ arbitrarily distributed gravitating masses $m_i$ and suppose periodic boundary conditions
for them on the boundary of the cell. Such models rely on background FLRW geometry with a scale factor $a$. It is supposed that the background solution is the
$\Lambda$CDM model with the perfect fluid in the form of dust with the average rest-mass density $\overline \rho$. Discrete inhomogeneities with the real
rest-mass density $\rho = \sum_{i=1}^N m_i\delta(\mathbf{r}-\mathbf{r}_i)$ perturb this background. The gravitational potential inside the cell is determined
in the Newtonian limit by the following Poisson equation \cite{Peebles,Bag,gadget2}:
%%%%%%%%
\be{1.1} \triangle\varphi (\mathbf{r})= 4\pi G_N \left[\sum_{i=1}^N m_i\delta(\mathbf{r}-\mathbf{r}_i)-\overline\rho\right]\, , \ee
%%%%%%%%
where $G_N$ is the Newtonian gravitational constant, and $\mathbf{r}, \mathbf{r}_i$ belong to the cell, e.g., $x_i\in [-l_1/2,l_1/2], y_i\in [-l_2/2,l_2/2],
z_i\in [-l_3/2,l_3/2]$. Here, the Laplace operator $\triangle =\partial^2/\partial x^2 +\partial^2/\partial y^2 + \partial^2/\partial z^2$, and the coordinates
$x,y$ and $z$, the gravitational potential $\varphi$ and the rest-mass densities $\rho$ and $\overline \rho$ correspond to the comoving frame. All these
quantities are connected with the corresponding physical ones as follows: $\mathbf{R}_{\mathrm{phys}} =a\mathbf{r}$, $\Phi_{\mathrm{phys}} = \varphi/a$ and
$\overline\rho_{\mathrm{phys}}=\overline\rho /a^3$. Eq. \rf{1.1} is the basic equation for the $N$-body simulation of the large scale structure formation in
the Universe \cite{gadget2}. The same equation can be also obtained in the Newtonian limit of General Relativity \cite{EZcosm2,EZcosm1,Durrer}. If we know the
gravitational potential, then we can investigate dynamics of the inhomogeneities/galaxies taking into account both gravitational attraction between them and
cosmological expansion of the Universe \cite{EZcosm1,EKZ2,EllGibb}.

It can be easily seen that in the case of a finite volume (e.g., the volume of the cell) Eq. \rf{1.1} satisfies the superposition principle. Here, for each
gravitating mass $m_i$ we can determine its contribution to the average rest-mass density: $\overline \rho_i=m_i/(l_1l_2l_3)$, $\overline \rho =
\sum_{i=1}^N\overline \rho_i$. Therefore, we can solve Eq. \rf{1.1} for each mass $m_i$ separately.

If we do not assume periodic boundary conditions, at least for one of directions, there is no lattice in these directions and space along them is not compact
(in the sense of lack of the finite period of the lattice).   Obviously, in infinite space the number of inhomogeneities must be also infinite: $ N \to
\infty$. This case has a number of potentially dangerous points. First, the superposition principle does not work here because we cannot determine $\overline
\rho_i$ for each of masses $m_i$. Second, it is known that the sum of an infinite number of Newtonian potentials diverges (the Neumann-Seeliger paradox
\cite{Norton}). Therefore, in general, the considered model can also suffer from this problem if we do not distribute masses in some specific way. Third, we
can easily see from Eq. \rf{1.1} that the presence of $\overline \rho$ will result in quadratic (with respect to the noncompact distance) divergence. Hence, to
avoid it, we should cut off gravitational potentials in these directions. This also may require a very specific distribution of the gravitating masses.

In the present paper, we investigate Eq. \rf{1.1} for different topologies of space which imply different kinds of the lattice structure. First, in section 2,
we consider the $T\times T\times T$ topology with periodic boundary conditions in all three spatial dimensions.  For point-like sources, we obtain a solution
in the form of an infinite series. This series has the well known Newtonian type divergence in the positions of masses. However, we show that the sum of the
series does not exist on the straight lines joining identical particles in neighboring cells. Therefore, there is no solution in points where masses are
absent. This is a new result. To avoid this nonphysical property, in section 3, we smear point-like sources. We present them in the form of uniformly filled
parallelepipeds. In this case, the infinite series has definite limits on the considered straight lines. Therefore, smearing of the gravitating masses in
$N$-body simulations plays a dual role: first, this is the absence of the Newtonian divergence in the positions of masses, second, this is the regular behavior
of the gravitational potential in all other points. Thus, in the present paper we provide a physical justification for such smearing.

In sections 4 and 5 we consider a possibility to get reasonable solutions of Eq. \rf{1.1} in the case of absence of periodicity in one or two spatial
directions. In section 4, we investigate a model with the spatial topology $T\times T\times R$,  i.e. with one noncompact dimension, let it be $z$. As we
mentioned above, due to noncompactness, the gravitational potential may suffer from the Neumann-Seeliger paradox and additionally has a divergence of the form
$\overline \rho z^2 \to +\infty $ for $|z|\to +\infty$. In this section we try to resolve these problems with the help of a special arrangement of gravitating
masses in the direction of $z$. Similar procedure in the flat Universe with topology $R^3$ was performed in \cite{EZcosm1}. Unfortunately, in the case of
topology $T\times T \times R$, there is no possibility to arrange masses in such a way that the gravitational potential is a smooth function in any point $z$.
The same result takes place for the Universe with topology $T\times R\times R$ which is considered in section 5. Here we also demonstrate impossibility of
constructing a smooth potential. The main results are briefly summarized in concluding section 6.

\section{Topology $T\times T\times T$. Point-like masses}

\setcounter{equation}{0}

Obviously, for topology $T\times T\times T$ the space is covered by identical cells, and, instead of an infinite number of these cells, we may consider just
one cell with periodic boundary conditions. As we mentioned in Introduction, due to the finite volume of the cell, we can apply the superposition principle. It
means that we can solve Eq. \rf{1.1} for one arbitrary gravitating mass, and the total gravitational potential in a point inside the cell is equal to a sum of
gravitational potentials (in this point) of all $N$ masses. Without loss of generality, we can put a gravitating mass $m$ at the origin of coordinates. Then,
the Poisson equation \rf{1.1} for this mass reads
%%%%%%%%%%%%%%%%%%%%
\be{2.1}
\Delta\varphi={4}\pi{G_N}\left({m}\delta\left({\bf
r}\right)-\frac{m}{l_{1}l_{2}l_{3}}\right)\, .
\ee
%%%%%%
Taking into account that delta-functions can be expressed as{\footnote{This is the standard Dirac delta-function representation for the considered geometry of
the model (see, e.g., \cite{BruLar1}).}}
%%%%%%
\be{2.2}
\delta(x)=\frac{1}{l_1} \sum_{k_1=-\infty}^{+\infty} \cos\left(\frac{2\pi k_1}{l_1}x\right)\, ,
\ee
%%%%%%
we get{\footnote{In the expression below, instead of the product $\cos(2\pi k_1 x/l_1) \cos(2\pi k_2 y/l_2) \cos(2\pi k_3 z/l_3)$ we can write $\cos(2\pi k_1
x/l_1+2\pi k_2 y/l_2+2\pi k_3 z/l_3)$, and with the help of the well-known formulas for the cosine of the sum this expression will give only the contribution
of the above mentioned form $\cos(2\pi k_1 x/l_1) \cos(2\pi k_2 y/l_2) \cos(2\pi k_3 z/l_3)$. Really, all terms containing, e.g., $\sin(2\pi k_1 x/l_1)$ (being
an odd function of $k_1$), will disappear from the sum with symmetric limits (i.e. $k_1$ varying from $-\infty$ to $+\infty$). }}
%%%%%%
\ba{2.3} \Delta\varphi&=&4\pi G_N \frac{m}{l_1 l_2 l_3}\left[\sum_{k_1=-\infty}^{+\infty}
\sum_{k_2=-\infty}^{+\infty}\sum_{k_3=-\infty}^{+\infty}\cos\left(\frac{2\pi k_1}{l_1}x\right)\right.\nn\\
&\times&\left.\cos\left(\frac{2\pi k_2}{l_2}y\right)\cos\left(\frac{2\pi k_3}{l_3}z\right)-1\right]\, . \ea
%%%%%%
Therefore, it makes sense to look for a solution of this equation in the form
%%%%%%%
\ba{2.4} \varphi&=&\sum_{k_1=-\infty}^{+\infty}\sum_{k_2=-\infty}^{+\infty}\sum_{k_3=-\infty}^{+\infty} C_{k_1 k_2 k_3}\nn\\
&\times&\cos\left(\frac{2\pi k_1}{l_1}x\right)\cos\left(\frac{2\pi k_2}{l_2}y\right)\cos\left(\frac{2\pi k_3}{l_3}z\right)\, , \ea
%%%%%%%
where unknown coefficients $C_{k_1 k_2 k_3}$ can be easily found from Eq. \rf{2.3}:
%%%%%%
\be{2.5}
C_{k_1 k_2 k_3}=-\frac{G_Nm}{\pi l_1 l_2 l_3} \frac{1}{\frac{k_1^2}{l_1^2}+\frac{ k_2^2}{l_2^2}+\frac{ k_3^2}{l_3^2}}\, , \quad
k_1^2+k_2^2+k_3^2\neq 0\, .
\ee
%%%%%%%
Hence, the desired gravitational potential is
%%%%%%%
\ba{2.6} \varphi&=&-\frac{G_Nm}{\pi l_1 l_2 l_3}\sum_{k_1=-\infty}^{+\infty} \sum_{k_2=-\infty}^{+\infty}\sum_{k_3=-\infty}^{+\infty}
\frac{1}{\frac{k_1^2}{l_1^2}+\frac{ k_2^2}{l_2^2}+\frac{ k_3^2}{l_3^2}}\nn\\
&\times&\cos\left(\frac{2\pi k_1}{l_1}x\right)\cos\left(\frac{2\pi k_2}{l_2}y\right)\cos\left(\frac{2\pi k_3}{l_3}z\right)\, , \ea
%%%%%%%
where $k_1^2+k_2^2+k_3^2\neq 0$. If $x,y,z$ simultaneously tend to zero, then the gravitational potential \rf{2.6} has the Newtonian limit:
%%%%%%%
\ba{2.7} &{}&\varphi\rightarrow-\frac{G_Nm}{\pi}\int\limits_{-\infty}^{\quad +\infty} dk_x\int\limits_{-\infty}^{\quad
+\infty}dk_y\int\limits_{-\infty}^{\quad +\infty} dk_z\nn\\
&{}&\frac{\cos(2\pi xk_x)\cos(2\pi yk_y)\cos(2\pi zk_z)}{k^2}=-\frac{G_Nm}{r}\, , \ea
%%%%%%%%
where $r=(x^2+y^2+z^2)^{1/2}$, as it should be. A good feature of the potential \rf{2.6} is that its average value (integral) over the cell is equal to zero:
$\overline \varphi =0$\footnote{It is worth noting that in \cite{EZcosm2,EZcosm1,EBV} the concrete mass distribution in the Universe with topology $R^3$ is
cited as an example of the case of nonzero average value $\overline \varphi\neq 0$. From the pure mathematical point of view this case is inadmissible in the
framework of the first-order perturbation theory.}. This is a physically reasonable result because $\overline{\rho -\overline{\rho}}=0$.

Clearly, in the case of a point-like gravitating source, we have usual divergence at the point of its location. Now, we want to demonstrate that there is also
a problem at points where gravitating masses are absent. More precisely, we will show that the sum \rf{2.6} is absent on straight lines which connect identical
masses in neighboring cells. In our particular example, they are lines of intersection (pairwise) of the planes $x=0$, $y=0$ and $z=0$. Let us consider the
potential \rf{2.6} on the straight line $y=0$, $z=0$. The numerical calculation of the potential on this straight line at the point $x=l_1/2$ for different
values of the limiting number $n$ (being the maximum absolute value of the summation indexes: $|k_{1,2,3}|\leq n$) is presented in the following table for the
cubic cell case $l_1=l_2=l_3\equiv l$. This table clearly demonstrates that the potential does not tend (with the growth of $n$) to any particular finite
number.

\begin{center}
\begin{tabular}{|c|c|c|c|}
\hline

\ & \ & \ & \

\\

{\Large $\ n\ $} & {\Large $\ \frac{\varphi_n(l/2,0,0)}{G_Nm/l}\ $} & {\Large $\ n\ $} &
{\Large $\ \frac{\varphi_n(l/2,0,0)}{G_Nm/l}\ $} \\

\ & \ & \ & \

\\

\hline

\ & \ & \ & \

\\

$40$ & $-0.73371$ & $41$ & $0.89453$ \\

%\ & \ & \ & \ \\

$60$ & $-0.72869$ & $61$ & $0.89969$ \\

%\ & \ & \ & \ \\

$80$ & $-0.72614$ & $81$ & $0.90229$ \\

\ & \ & \ & \

\\

\hline
\end{tabular}
\end{center}

To understand the reason for this, let us analyze the structure of the expression \rf{2.6} in more detail. For $z=0$ the gravitational potential reads:
%%%%%
\ba{2.8} &{}&\varphi (x,y,0)=-\frac{G_Nm}{\pi l_1 l_2 l_3}\nn\\
&\times&\sum_{k_1=-\infty}^{+\infty} \sum_{k_2=-\infty}^{+\infty}\sum_{k_3=-\infty}^{+\infty} \frac{\cos\left(\frac{2\pi
k_1}{l_1}x\right)\cos\left(\frac{2\pi k_2}{l_2}y\right)}{\frac{k_1^2}{l_1^2}+\frac{ k_2^2}{l_2^2}+\frac{ k_3^2}{l_3^2}}\nn\\
&=& -\frac{G_Nm \pi
l_3}{3l_1 l_2}-\frac{4G_Nm}{l_1 l_2 } \sum_{k_1=1}^{+\infty} \sum_{k_2=1}^{+\infty} \frac{1}{\sqrt{\frac{k_1^2}{l_1^2}+\frac{k_2^2}{l_2^2}}} \nn\\
&\times&\cos\left(\frac{2\pi k_1}{l_1}x\right)\cos\left(\frac{2\pi k_2}{l_2}y\right)
\coth\left(\pi\, \sqrt{\frac{l_3^2k_1^2}{l_1^2}+\frac{ l_3^2k_2^2}{l_2^2}}\right)\nn \\
&-& \frac{2G_Nm}{ l_2 }\sum_{k_1=1}^{+\infty}\frac{1}{k_1} \cos\left(\frac{2\pi k_1}{l_1}x\right)
\coth\left(\frac{\pi l_3k_1}{l_1}\right)\nn\\
&-& \frac{2G_Nm}{ l_1} \sum_{k_2=1}^{+\infty}\frac{1}{k_2} \cos\left(\frac{2\pi k_2}{l_2}y\right) \coth\left(\frac{\pi l_3k_2}{l_2}\right)\, , \ea
%%%%%%
where we used the tabulated formulas for sums of series \cite{Prud} (see, e.g., 5.1.25). All sums in this expression are potentially dangerous. To show it, we
can drop the hyperbolic cotangents because $\coth k \to 1$ for $k\to +\infty$. Two last sums are divergent depending on which straight line we consider: $x=0$
or $y=0$, respectively. For example, on the straight line $y=0$, the sum $\sum_{k_1=1}^{+\infty} \cos\left(2\pi k_1x/l_1\right)/k_1 = -\ln \left[2 \sin(\pi
x/l_1)\right]$ (see 5.4.2 in \cite{Prud}) is convergent for any ratio $x/l_1 \neq 0,1$ while $\sum_{k_2=1}^{+\infty}(1/k_2)\sim \lim\limits_{k_2\to +\infty}\ln
k_2$ is logarithmically divergent. The rough estimate of the double sum also leads to a divergent result.
%: $\sum_{k_1}\sum_{k_2}\cos(k_1x)/\sqrt{k_1^2+k_2^2}\sim\sum_{k_1}\cos(k_1x) \ln k_1$
To be more precise, we investigate now finite sums\footnote{Obviously, the inclusion of the hyperbolic cotangents does not effect the main results but makes
calculations more complicated.} of the "suspicious" terms on the straight line $y=0$:
%%%%%%%
\be{2.12}
f_n(x) = 2\sum_{k_1=1}^{n} \sum_{k_2=1}^{n}\cos\left(\frac{2\pi
k_1}{l_1}x\right)
\frac{1}{\sqrt{\frac{l_2^2}{l_1^2}k_1^2+k_2^2}}
+ \sum_{k_2=1}^{n} \frac{1}{k_2}\, .
\ee
%%%%%%%
It is worth noting that in the case $l_1=l_2$ and $x/l_1=1/2$ the logarithmically divergent terms exactly cancel each other. Really, it follows directly from
the following estimates:
%%%%%%%
\ba{2.12a}
&{}&2\sum_{k_1=1}^{+\infty} \sum_{k_2=1}^{+\infty}\frac{\cos\left(\pi
k_1\right)}
{\sqrt{k_1^2+k_2^2}} = 2\sum_{k_1=1}^{+\infty} \sum_{k_2=1}^{+\infty}\frac{\left(-1\right)^{k_1}}
{\sqrt{k_1^2+k_2^2}}\nn\\
&{=}& 2\sum_{m=1}^{+\infty} \sum_{k_2=1}^{+\infty}\left[\frac{1}
{\sqrt{(2m)^2+k_2^2}} -\frac{1}{\sqrt{(2m-1)^2+k_2^2}}\right]\nn \\
&\sim&- 2\sum_{m=1}^{+\infty} \sum_{k_2=1}^{+\infty}\frac{2m} {\left[(2m)^2+k_2^2\right]^{3/2}}\nn\\
&\sim& -2\int\limits_1^{+\infty}\int\limits_1^{+\infty} \frac{2x}{\left[(2x)^2+y^2\right]^{3/2}}dxdy\nn\\
&\sim& - \lim\limits_{R\to +\infty}\ln R =-\infty
\ea
%%%%%%%
and
%%%%%%%
\be{2.12b}
\sum_{k_2=1}^{+\infty}\frac{1}{k_2}\sim\int\limits_{1}^{+\infty}\frac{dx}{x} \sim \lim \limits_{R\to +\infty}\ln R =+\infty\, .
\ee
%%%%%%%
Therefore, both of these logarithmically divergent terms cancel each other. Nevertheless, the expression \rf{2.12} does not have a definite limit for
$n\to+\infty$. To demonstrate it, along with \rf{2.12} let us introduce the function
%%%%%%%
\ba{2.13} f_{n+1}(x) &=& 2\sum_{k_1=1}^{n+1} \sum_{k_2=1}^{n+1}\cos\left(\frac{2\pi k_1}{l_1}x\right)\nn\\
&\times& \frac{1}{\sqrt{\frac{l_2^2}{l_1^2}k_1^2+k_2^2}} + \sum_{k_2=1}^{n+1} \frac{1}{k_2}\, . \ea
%%%%%%%
Evidently, if the expression \rf{2.12} is convergent for $n\to +\infty$, then in this limit the difference $f_{n+1}(x)-f_{n}(x) \to 0$. After some simple
algebra we get (for $l_1=l_2$)
%%%%%%%
\ba{2.14} &{}&f_{n+1}(x)-f_{n}(x)\nn\\
&=& 2\sum_{k_1=1}^{n} \cos\left(\frac{2\pi k_1}{l_1}x\right)\frac{1}{\sqrt{k_1^2+(n+1)^2}}\nn \\
&+& 2\sum_{k_2=1}^{n} \cos\left(\frac{2\pi (n+1)}{l_1}x\right)\frac{1}{\sqrt{(n+1)^2+k_2^2}}\nn\\
&+& \frac{1+\sqrt{2}\cos \left(2\pi (n+1)x/l_1\right)}{n+1}\nn \\
&\equiv& \triangle f_n(x) + \frac{1+\sqrt{2}\cos \left(2\pi (n+1)x/l_1\right)}{n+1}\, . \ea
%%%%%%
Here, the last term in the third line vanishes for $n\to +\infty$. Therefore, the problem of convergence of \rf{2.12} is reduced now to the analysis of
$\triangle f_n(x)$. In Figure 1, we show the graph of $\triangle f_n(x)$ (for $x/l_1=1/2$) as a function of $n$. Each point gives the value of $\triangle
f_n(x)$ for the corresponding number $n$. This picture clearly demonstrates that the difference $f_{n+1}(x)-f_{n}(x)$ does not tend to zero for growing $n$.
Even more, it does not go to any definite value.
%(jumping from one value to another with changing $n$).
It can be also verified that a similar result takes place for any other point on any of the straight lines and holds also for $\l_1\neq l_2$. Therefore, we
have proven that in the case of point-like gravitating masses in the considered lattice Universe the gravitational potential has no definite values on the
straight lines joining identical masses in neighboring cells. Clearly, this is a nonphysical result since the dynamics of cosmic bodies is not determined in
such a case.

\

\begin{figure}[hbt]
\center{\includegraphics[width=8.5cm,height=5cm]{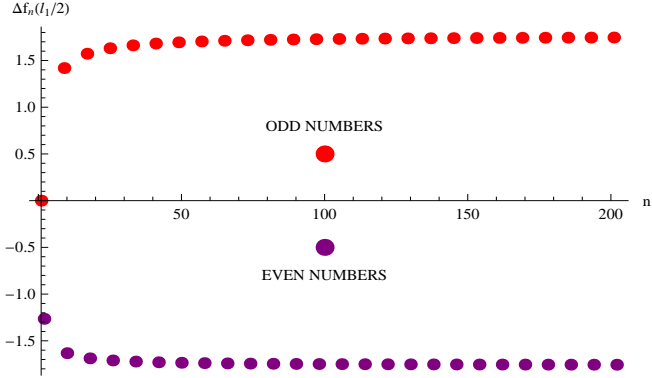}} \caption{The graph of $\triangle f_n(l_1/2)$ as a function of the number $n$.}
\end{figure}

%%%%%%%%%%%%%%%%%%%%%%%%%%%%%%%%%%%%%%%%%%%%%%%%%%%%%%%%%%%%%%%%%%%%%%%%%%%%%%%%%%%%%%%%%%%%%%%%%%%%%%%
%%%%%%%%%%%%%%%%%%%%%%%%%%%%%%%%%%%%%%%%%%%%%%%%%%%%%%%%%%%%%%%%%%%%%%%%%%%%%%%%%%%%%%%%%%%%%%%%%%%%%%%

\

\section{Topology $T\times T\times T$. Smeared masses}

\setcounter{equation}{0}

Can the smearing of gravitating masses resolve the problem found in the previous section? To answer this question, we present gravitating masses as uniformly
filled parallelepipeds. This representation of masses looks a bit artificial. However, such form is the most appropriate for the considered cells, and the most
important point is that the form of smearing does not matter for us at the moment. We just want to get a principal answer on a possibility to avoid the problem
with the help of smearing. So, let the mass $m$ be uniformly smeared over a parallelepiped (with the lengthes of edges $a,b$ and $c$) which we put, without
loss of generality, in the middle of the cell. It is convenient to introduce a function $f_1(x)$ equal to $1$ for $x\in [-a/2,a/2]$ and $0$ elsewhere inside
$[-l_1/2,l_1/2]$. We can write down this function as
%%%%%%%
\be{3.1} f_1(x)=\frac{a}{l_1}+\sum_{n=1}^{+\infty} \frac{2}{\pi n}\sin\left(\frac{a\pi n}{l_1}\right)\cos\left(\frac{2\pi n}{l_1}x\right)\, . \ee
%%%%%%%
Similarly,
%%%%%%
\be{3.2} f_2(y)=\frac{b}{l_2}+\sum_{j=1}^{+\infty} \frac{2}{\pi j}\sin\left(\frac{b\pi j}{l_2}\right)\cos\left(\frac{2\pi j}{l_2}y\right)\, \ee
%%%%%%
and
%%%%%%%
\be{3.3} f_3(z)=\frac{c}{l_3}+\sum_{k=1}^{+\infty} \frac{2}{\pi k}\sin\left(\frac{c\pi k}{l_3}\right)\cos\left(\frac{2\pi k}{l_3}z\right)\, . \ee
%%%%%%%%
Therefore, the rest-mass density of the mass under consideration is
%%%%%%%
\be{3.4}
\rho (\mathbf{r}) = \frac{m}{abc}f_1(x)f_2(y)f_3(z) \equiv \frac{m}{abc}f(\mathbf{r})\, .
\ee
%%%%%%%
Then, Eq. \rf{1.1} for this mass reads:
%%%%%%
\ba{3.5} \triangle\varphi &=& 4\pi G_N\left[\frac{m}{abc}f({\bf r})-\frac{m}{l_1l_2l_3}\right]\nn\\
&=&4\pi G_Nm\left[\frac{1}{l_1l_2c}\sum_{k=1}^{+\infty} \frac{2}{\pi
k}\sin\left(\frac{c\pi k}{l_3}\right)\cos\left(\frac{2\pi k}{l_3}z\right)\right.\nn\\
&+&\frac{1}{l_1l_3b}\sum_{j=1}^{+\infty} \frac{2}{\pi j}\sin\left(\frac{b\pi j}{l_2}\right)\cos\left(\frac{2\pi
j}{l_2}y\right)\nn\\
&+&\frac{1}{l_2l_3a}\sum_{n=1}^{+\infty}
\frac{2}{\pi n}\sin\left(\frac{a\pi n}{l_1}\right)\cos\left(\frac{2\pi n}{l_1}x\right)\nn \\
&+&\frac{1}{l_1bc}\sum_{j=1}^{+\infty}\sum_{k=1}^{+\infty} \frac{4}{\pi^2jk}\sin\left(\frac{b\pi j}{l_2}\right)
\cos\left(\frac{2\pi j}{l_2}y\right)\nn\\
&\times&\sin\left(\frac{c\pi k}{l_3}\right)\cos\left(\frac{2\pi k}{l_3}z\right)\nn \\
&+&\frac{1}{l_2ac}\sum_{n=1}^{+\infty}\sum_{k=1}^{+\infty} \frac{4}{\pi^2nk}\sin\left(\frac{a\pi n}{l_1}\right)\cos\left(\frac{2\pi
n}{l_1}x\right)\nn\\
&\times&\sin\left(\frac{c\pi
k}{l_3}\right)\cos\left(\frac{2\pi k}{l_3}z\right)\nn \\
&+&\frac{1}{l_3ab}\sum_{n=1}^{+\infty}\sum_{j=1}^{+\infty} \frac{4}{\pi^2 nj}\sin\left(\frac{a\pi n}{l_1}\right)\cos\left(\frac{2\pi
n}{l_1}x\right)\nn\\
&\times&\sin\left(\frac{b\pi j}{l_2}\right)\cos\left(\frac{2\pi j}{l_2}y\right)\nn\\
&+&\frac{1}{abc}\sum_{n=1}^{+\infty}\sum_{j=1}^{+\infty}\sum_{k=1}^{+\infty} \frac{8}{\pi^3njk}\nn\\
&\times&\sin\left(\frac{a\pi n}{l_1}\right)\sin\left(\frac{b\pi
j}{l_2}\right)\sin\left(\frac{c\pi k}{l_3}\right)\nn \\
&\times&\left.\cos\left(\frac{2\pi n}{l_1}x\right)\cos\left(\frac{2\pi j}{l_2}y\right)\cos\left(\frac{2\pi k}{l_3}z\right)\right]\, . \ea
%%%%%%%
This equation implies that it makes sense to look for a solution in the following form:
\ba{3.6}
\varphi (\mathbf{r})&=&\frac{m}{l_1l_2c}\sum_{k=1}^{+\infty} C_{k}\sin\left(\frac{c\pi k}{l_3}\right)\cos\left(\frac{2\pi k}{l_3}z\right)\nn \\
&+&\frac{m}{l_1l_3b}\sum_{j=1}^{+\infty} C'_{j}\sin\left(\frac{b\pi j}{l_2}\right)\cos\left(\frac{2\pi
j}{l_2}y\right)\nn\\
&+&\frac{m}{l_2l_3a}\sum_{n=1}^{+\infty}
C''_{n}\sin\left(\frac{a\pi n}{l_1}\right)\cos\left(\frac{2\pi n}{l_1}x\right)\nn\\
&+&\frac{m}{l_1bc}\sum_{j=1}^{+\infty}\sum_{k=1}^{+\infty} C_{jk}\sin\left(\frac{b\pi j}{l_2}\right)\cos\left(\frac{2\pi
j}{l_2}y\right)\nn\\
&\times&\sin\left(\frac{c\pi
k}{l_3}\right)\cos\left(\frac{2\pi k}{l_3}z\right)\nn \\
&+&\frac{m}{l_2ac}\sum_{n=1}^{+\infty}\sum_{k=1}^{+\infty} C'_{nk}\sin\left(\frac{a\pi n}{l_1}\right)\cos\left(\frac{2\pi
n}{l_1}x\right)\nn\\
&\times&\sin\left(\frac{c\pi
k}{l_3}\right)\cos\left(\frac{2\pi k}{l_3}z\right)\nn \\
&+&\frac{m}{l_3ab}\sum_{n=1}^{+\infty}\sum_{j=1}^{+\infty} C''_{nj}\sin\left(\frac{a\pi n}{l_1}\right)\cos\left(\frac{2\pi
n}{l_1}x\right)\nn\\
&\times&\sin\left(\frac{b\pi
j}{l_2}\right)\cos\left(\frac{2\pi j}{l_2}y\right)\nn \\
&+&\frac{m}{abc}\sum_{n=1}^{+\infty}\sum_{j=1}^{+\infty}\sum_{k=1}^{+\infty} C_{njk}\nn\\
&\times&\sin\left(\frac{a\pi n}{l_1}\right)\sin\left(\frac{b\pi
j}{l_2}\right)\sin\left(\frac{c\pi k}{l_3}\right)\nn \\
&\times&\cos\left(\frac{2\pi n}{l_1}x\right)\cos\left(\frac{2\pi j}{l_2}y\right)\cos\left(\frac{2\pi k}{l_3}z\right)\, . \ea
%%%%%%%
Substitution of this expression into the Poisson equation \rf{3.5} gives
%%%%%
\ba{3.7} C_{k}&=&-\frac{2G_N}{\pi^2k}\left(\frac{l_3}{k}\right)^2\, ,\ \ C'_{j}=-\frac{2G_N}{\pi^2j}\left(\frac{l_2}{j}\right)^2\, ,\ \ \nn\\
C''_{n}&=&-\frac{2G_N}{\pi^2n}\left(\frac{l_1}{n}\right)^2\, ,\nn \\
C_{jk}&=&-\frac{4G_N}{\pi^3jk}\frac{1}{\left(\frac{j^2}{l_2^2}+\frac{k^2}{l_3^2}\right)}\, ,\ \
C'_{nk}=-\frac{4G_N}{\pi^3nk}\frac{1}{\left(\frac{n^2}{l_1^2}+\frac{k^2}{l_3^2}\right)}\, ,\nn\\
C''_{jn}&=&-\frac{4G_N}{\pi^3nj}\frac{1}{\left(\frac{n^2}{l_1^2}+\frac{j^2}{l_2^2}\right)}\, ,\ \ \nn\\
C_{njk}&=&-\frac{8G_N}{\pi^4njk}\frac{1}{\left(\frac{n^2}{l_1^2}+\frac{j^2}{l_2^2}+\frac{k^2}{l_3^2}\right)}\, . \ea
%%%%%%%
Let us choose the same straight line as in the previous section, that is $y=0$, $z=0$, and the same point $x=l_1/2$. The numerical calculation of the
gravitational potential in this point for different values of the limiting number $n$ is presented in the following table for the cubic cell case under the
additional condition $a=b=c=(3/7)l$. In contrast to the previous case of a point-like source, here the potential apparently tends to a particular finite value.
Therefore, in the case of smeared gravitating masses the gravitational potential has a regular behavior at any point inside the cell (including, e.g., the
point $x=y=z=0$).

\begin{center}
\begin{tabular}{|c|c|c|c|}
\hline

\ & \ & \ & \

\\

{\Large $\ n\ $} & {\Large $\ \frac{\varphi_n(l/2,0,0)}{G_Nm/l}\ $} & {\Large $\ n\ $} &
{\Large $\ \frac{\varphi_n(l/2,0,0)}{G_Nm/l}\ $} \\

\ & \ & \ & \

\\

\hline

\ & \ & \ & \

\\

$15$ & $0.028717$ & $16$ & $0.028443$ \\

%\ & \ & \ & \ \\

$19$ & $0.028536$ & $20$ & $0.028222$ \\

%\ & \ & \ & \ \\

$23$ & $0.028368$ & $24$ & $0.028223$ \\

\ & \ & \ & \

\\

\hline
\end{tabular}
\end{center}

%%%%%%%%%%%%%%%%%%%%%%%%%%%%%%%%%%%%%%%%%%%%%%%%%%%%%%%%%%%%%%%%%%%%%%%%%%%%%%%%%%%%
%%%%%%%%%%%%%%%%%%%%%%%%%%%%%%%%%%%%%%%%%%%%%%%%%%%%%%%%%%%%%%%%%%%%%%%%%%%%%%%%%%%%

\section{Topology $T\times T\times R$}

\setcounter{equation}{0}

The $T\times T\times R$ topology implies one noncompact dimension, let it be $z$. Therefore, there is a lattice structure in directions $x$ and $y$ and an
irregular structure in the direction $z$. In a column $x\in [-l_1/2,l_1/2]$, $y\in [-l_2/2,l_2/2]$, $z\in (-\infty,+\infty)$ there is an infinite number of
gravitating masses. To obtain a "nice" regular solution, we will try to arrange masses in the $z$ direction in such a way that in each point $z$ the
gravitational potential is determined by one mass only. There are two possibilities for that. Let this mass be at $z=0$. In the first scenario, the potential
and its first derivative (with respect to $z$) should vanish at some distance $z_0$ (which we determine below). Then, the next mass should be at a distance (in
the $z$ direction) equal or greater than $z_0+z_1$, where $z_1$ is a distance at which the gravitational potential and its first derivative vanish for the
second mass. Similarly, we should shift in the direction of $z$ the third mass with respect to the second one and so on. In this scenario, we can arrange
strips $\Delta z$ between masses where the potential is absent. It occurs, e.g., between the first and second masses if the second mass is situated at
distances greater that $z_0+z_1$. In the strip, we place a uniform medium with the rest-mass density $\overline \rho$. Coordinates $x\in [-l_1/2,l_1/2]$ and
$y\in [-l_2/2,l_2/2]$ of masses are arbitrary. In the second scenario, we should determine distances $z_0$, $z_1$, $z_2,\ \ldots$ where potentials of
neighboring (in the $z$ direction) particles are smoothly matched to each other. This means that at these distances potentials are generally nonzero. Moreover,
we suppose that their first derivatives are zero at the points of matching, i.e. potentials have extrema in these points. In this scenario, the neighboring (in
the $z$ direction) masses should have the same coordinates $x$ and $y$. Now let us consider these scenarios in detail. For both of them, we need to look for a
solution just for one particle. Let this particle be in the point $x=y=z=0$. Then, Eq. \rf{1.1} reads
%%%%%%
\be{4.1}
\triangle\varphi=4\pi G_N\left(m\delta({\bf r})-\bar\rho\right)\, .
\ee
%%%%%%
Keeping in mind the regular structure in $x$ and $y$ directions, we can represent the delta functions $\delta (x)$ and $\delta (y)$ in the form \rf{2.2}. So,
Eq. \rf{4.1} is reduced to
%%%%%%
\ba{4.2} \triangle\varphi &=& 4\pi G_N\left[\frac{m}{l_1l_2}\sum_{k_1=-\infty}^{+\infty}\sum_{k_2=-\infty}^{+\infty}\cos\left(\frac{2\pi
k_1}{l_1}x\right)\right.\nn\\
&\times&\left.\cos\left(\frac{2\pi k_2}{l_2}y\right)\delta(z)-\bar\rho\right]\, . \ea
%%%%%%
Evidently, we can look for a solution of this equation in the form
%%%%%%
\be{4.3} \varphi=\sum_{k_1=-\infty}^{+\infty}\sum_{k_2=-\infty}^{+\infty}C_{k_1k_2}(z)\cos\left(\frac{2\pi k_1}{l_1}x\right)\cos\left(\frac{2\pi
k_2}{l_2}y\right)\, , \ee
%%%%%%
and from the Poisson equation \rf{4.2} we get
%%%%%%
\ba{4.4} &{}&G_N\bar\rho=\sum_{k_1=-\infty}^{+\infty}\sum_{k_2=-\infty}^{+\infty}\pi\cos\left(\frac{2\pi k_1}{l_1}x\right)\cos\left(\frac{2\pi
k_2}{l_2}y\right)\nn\\
&\times&\left[\left(\frac{k_2^2}{l_2^2}+\frac{k_1^2}{l_1^2}\right)C_{k_1k_2}(z)+\frac{mG_N}{l_1l_2\pi}\delta(z)-\frac{C_{k_1k_2}^{''}(z)}{4\pi^2}\right]\, .\nn\\
&{}&\ea
%%%%%%
In this section, the prime denotes the derivative with respect to $z$. Now, we should determine unknown functions $C_{k_1k_2}(z)$. First, we find the zero mode
$C_{00}(z)$ which satisfies the equation
%%%%%%
\be{4.5}
\frac{C_{00}^{''}(z)}{4\pi}=\frac{mG_N}{l_1l_2}\delta(z)-G_N\bar\rho\, .
\ee
%%%%%%
This equation has the solution
%%%%%%
\be{4.6}
C_{00}(z)=-2\pi G_N\bar\rho z^2+\frac{2\pi}{l_1l_2}mG_N|z|+B\, ,
\ee
%%%%%%
where $B$ is a constant of integration. This solution is a function growing with $z$. Therefore, we must cut off it at some distance $z_0$.

Let us consider the first scenario. From the condition $C'_{00}(z_0)=0$ we obtain
%%%%%%
\be{4.7}
z_0=\frac{m}{2\bar\rho l_1l_2}\quad \Leftrightarrow\quad \bar\rho=\frac{m}{2 z_0 l_1l_2}\, .
\ee
%%%%%%
The second condition $C_{00}(z_0)=0$ provides the value of $B$:
%%%%%%
\be{4.8}
B=-\frac{\pi m^2G_N}{2\bar\rho l_1^2l_2^2}\, .
\ee
%%%%%%
Now, we want to determine the form of $C_{k_1k_2}(z)$ when $k_1^2+k_2^2 \neq 0$. In this case Eq. \rf{4.4} is consistent only if the following condition holds
true:
%%%%%%
\ba{4.9} &{}&\left(\frac{k_2^2}{l_2^2}+\frac{k_1^2}{l_1^2}\right)C_{k_1k_2}(z)+\frac{mG_N}{l_1l_2\pi}\delta(z)-\frac{C_{k_1k_2}^{''}(z)}{4\pi^2}=0\, ,\nn\\
&{}&k_1^2+k_2^2 \neq 0\, . \ea
%%%%%%
We look for a solution of this equation in the form
%%%%%%
\be{4.10}
C_{k_1k_2}(z)=\widetilde{A}e^{-2\pi\beta |z|} + \widetilde{B}e^{2\pi\beta |z|}\, ,\quad \beta\equiv \sqrt{\frac{k_2^2}{l_2^2}+\frac{k_1^2}{l_1^2}}\, ,
\ee
%%%%%%
where $\widetilde{A}$ and $\widetilde{B}$ are constants. The substitution of this function into Eq. \rf{4.9} gives
%%%%%%
\be{4.11}
\widetilde{A} -\widetilde{B} = - \frac{m G_N}{l_1l_2\beta}\, .
\ee
%%%%%%
Therefore,
%%%%%%
\ba{4.12} &{}&C_{k_1k_2}(z)= 2 \widetilde{B}\cosh (2\pi \beta z) - \frac{m G_N}{l_1l_2\beta}e^{-2\pi\beta |z|}\, ,\nn\\
&{}& k_1^2+k_2^2 \neq 0\, . \ea
%%%%%%
From the boundary condition $C_{k_1k_2}(z_0)=0$ we get
%%%%%%
\be{4.13} \widetilde{B} = \frac{m G_N}{2l_1l_2\beta} e^{-2\pi\beta z_0}\left[\cosh (2\pi \beta z_0)\right]^{-1}\, . \ee
%%%%%%
It can be easily verified that the function \rf{4.12} (with $\widetilde{B}$ from \rf{4.13}) does not satisfy the boundary condition $C'_{k_1k_2}(z_0)=0$.
Hence, we cannot determine the gravitational potential in accordance with the first scenario.

Now, we intend to demonstrate that there is a possibility to find the potential in the framework of the second scenario in the case of identical masses.
However, this construction has a drawback inherent in the $T\times T\times T$ model with the point-like source.

In the second scenario with identical masses $m$, all of them have the same coordinates $x,y$ and are separated by the same distance $2z_0\equiv l_3$ in the
direction of $z$. Here, the function $C_{00}(z)$ still has the form \rf{4.6}. Since we require $C'_{00}(z_0)=0$, the boundary $z_0$ is determined by \rf{4.7}.
However, the constant $B$ is not given now by \rf{4.8} because the condition $C_{00}(z_0)=0$ is absent. This constant can by found from the condition
$\overline\varphi=0 $ over the period $l_3=2z_0$. That is
%%%%%%
\ba{4.14} \int_{-z_0}^{+z_0}C_{00}(z)dz&=&0 \quad\Rightarrow \nn\\
B &=& - \frac{2\pi G_N m z_0}{3l_1l_2} =-\frac{\pi G_N m l_3}{3l_1l_2}\, . \ea
%%%%%
The functions $C_{k_1k_2}(z)$ for $k_1^2+k_2^2\neq 0$ are given by Eq. \rf{4.12}, where the constant $\widetilde B$ follows from the boundary condition
$C'_{k_1k_2}(z_0)=0$:
%%%%%%
\be{4.15} \widetilde B = -\frac{m G_N}{2l_1l_2\beta} e^{-2\pi\beta z_0}\left[\sinh (2\pi \beta z_0)\right]^{-1}\, . \ee
%%%%%%
It can be easily verified that $C_{k_1k_2}(z)$ can be rewritten in the form
%%%%%%
\be{4.16} C_{k_1k_2}(z)= - \frac{ G_N m}{l_1l_2\beta\sinh(2\pi\beta z_0)}\cosh\left[2\pi\beta(|z|-z_0)\right]\, . \ee
%%%%%%
Therefore, in the second scenario the gravitational potential is
%%%%%%
\ba{4.17}
\varphi&=&\sum_{k_1=-\infty}^{+\infty}\sum_{k_2=-\infty}^{+\infty}C_{k_1k_2}(z)\cos\left(\frac{2\pi k_1}{l_1}x\right)\cos\left(\frac{2\pi k_2}{l_2}y\right)\nn \\
&=&C_{00}(z)+2\sum_{k_1=1}^{+\infty}C_{k_10}(z)\cos\left(\frac{2\pi k_1}{l_1}x\right)\nn\\
&+&2\sum_{k_2=1}^{+\infty}C_{0k_2}(z)\cos\left(\frac{2\pi k_2}{l_2}y\right)\nn \\
&+&4\sum_{k_1=1}^{+\infty}\sum_{k_2=1}^{+\infty}C_{k_1k_2}(z)\cos\left(\frac{2\pi k_1}{l_1}x\right)\cos\left(\frac{2\pi k_2}{l_2}y\right)\nn\\
&=& -G_N m\left\{\frac{2\pi}{l_1l_2l_3} z^2-\frac{2\pi}{l_1l_2}|z|+\frac{\pi l_3}{3l_1l_2}\right.\nn \\
&+&\frac{2}{l_2}\sum_{k_1=1}^{+\infty}\frac{\cos\left(2\pi k_1 x/l_1\right)}{k_1}\; \frac{\cosh[2\pi k_1(|z|-z_0)/l_1]}{\sinh(2\pi k_1 z_0/l_1)} \nn \\
&+&
\frac{2}{l_1}\sum_{k_2=1}^{+\infty}\frac{\cos\left(2\pi k_2 y/l_2\right)}{k_2}\; \frac{\cosh[2\pi k_2(|z|-z_0)/l_2]}{\sinh(2\pi k_2 z_0/l_2)} \nn\\
&+&\left.\frac{4}{l_1l_2}\sum_{k_1=1}^{+\infty}\sum_{k_2=1}^{+\infty}\frac{\cos\left(2\pi k_1 x/l_1\right) \cos\left(2\pi k_2 y/l_2\right)}
{\sqrt{\frac{k_1^2}{l_1^2}+\frac{k_2^2}{l_2^2}}}\right.\nn\\
&\times&\left.\frac{\cosh\left[2\pi
\sqrt{\frac{k_1^2}{l_1^2}+\frac{k_2^2}{l_2^2}}(|z|-z_0)\right]}{\sinh\left(2\pi\sqrt{\frac{k_1^2}{l_1^2}+\frac{k_2^2}{l_2^2}} z_0\right)}\right\}\, . \ea
%%%%%%
When $z=0$ and $x,y$ simultaneously go to zero, the potential $\varphi\rightarrow-G_Nm/\sqrt{x^2+y^2}$, as it should be. From the physical point of view, it is
clear that this scenario should coincide with the $T\times T\times T$ case. Really, the triple sum \rf{2.6} can be rewritten in the form \rf{4.17} with the
help of \cite{Prud} (see 5.4.5). It can be also easily seen that on the plane $z=0$ the expression \rf{4.17} exactly coincides with Eq. \rf{2.8}. Therefore, in
the second scenario we again arrive at the nonphysical result that the gravitational potential has no definite values on straight lines $y=0$ and $x=0$.

%%%%%%%%%%%%%%%%%%%%%%%%%%%%%%%%%%%%%%%%%%%%%%%%%%%%%%%%%%%%%%%%%%%%%%%%%%%%%%%
%%%%%%%%%%%%%%%%%%%%%%%%%%%%%%%%%%%%%%%%%%%%%%%%%%%%%%%%%%%%%%%%%%%%%%%%%%%%%%%

\section{Topology $T\times R\times R$}

\setcounter{equation}{0}

In this section we consider a model with a periodic boundary condition in one direction only. Two other spatial dimensions are noncompact. Here, in analogy
with the previous section, we also suppose that the gravitational potential in the vicinity of a particle is determined by its mass only. The shape of such
domain is dictated by the symmetry of the model and will be described below. On the boundary (in the direction of noncompact dimensions) of this domain the
potential and its first derivative are equal to zero, and between such domains the potential is absent: $\varphi =0$. Therefore, this model is similar to the
first scenario in the previous section.

Let the mass be at the point $x=y=z=0$ and the periodic (with the period $l_1$) boundary condition be along the $x$ coordinate. Then, the Poisson equation
\rf{4.1} for this topology can be written as follows:
%%%%%%
\be{5.1} \triangle\varphi =4\pi G_N\frac{m}{l_1}\sum_{k_1=-\infty}^{+\infty}\cos\left(\frac{2\pi k_1}{l_1}x\right)\delta(y)\delta(z)-4\pi G_N\bar\rho\, . \ee
%%%%%%
We look for a solution of this equation in the form
%%%%%%
\be{5.2} \varphi=\sum_{k_1=-\infty}^{+\infty}C_{k_1}(y,z)\cos\left(\frac{2\pi k_1}{l_1}x\right)\, . \ee
%%%%%%%
Then, from the Poisson equation \rf{5.1} we get
%%%%%%
\ba{5.3} G_N\bar\rho&=&\sum_{k_1=-\infty}^{+\infty}\cos\left(\frac{2\pi k_1}{l_1}x\right)\nn\\
&\times&\left[G_N\frac{m}{l_1}\delta(y)\delta(z)+ \frac{\pi k_1^2}{l_1^2}C_{k_1}(y,z)\right.\nn\\
&-&\left.\frac{C''_{k_1y}(y,z)}{4\pi}-\frac{C''_{k_1z}(y,z)}{4\pi}\right]\, , \ea
%%%%%%
where
%%%%%%
\be{5.4}
C''_{k_1y}(y,z) \equiv \frac{\partial^2 C_{k_1}(y,z)}{\partial y^2}\, , \quad
C''_{k_1z}(y,z) \equiv \frac{\partial^2 C_{k_1}(y,z)}{\partial z^2}\, .
\ee
%%%%%%
For the zero mode $k_1=0$ this equation gives
%%%%%%
\be{5.5}
G_N\bar\rho=G_N\frac{m}{l_1}\delta(y)\delta(z)-\frac{C''_{0y}(y,z)}{4\pi}-\frac{C''_{0z}(y,z)}{4\pi}\, .
\ee
%%%%%%
Following the geometry of the model, it makes sense to turn to polar coordinates:
%%%%%%
\be{5.6}
y=\xi\cos\phi, \quad z=\xi\sin\phi\, .
\ee
%%%%%%
Then, the two-dimensional Laplace operator is
%%%%%%
\be{5.7} \triangle_{\xi\phi}=\frac{1}{\xi}\frac{\partial}{\partial \xi}\left(\xi\frac{\partial}{\partial \xi}\right)+
\frac{1}{\xi^2}\frac{\partial^2}{\partial\phi^2} \equiv \triangle_{\xi} + \triangle_{\phi}\, . \ee
%%%%%%
It is clear that due to the symmetry of the problem the functions $C_{k_1}$ do not depend on the azimuthal angle $\phi$. Therefore, Eq. \rf{5.5} reads
%%%%%%
\be{5.8} G_N\bar\rho=G_N\frac{m}{l_1}\delta\left(\vec{\xi}\, \right)-\frac{\triangle_\xi C_0}{4\pi}\, . \ee
%%%%%%
This equation has the solution
%%%%%%
\be{5.9} C_0=-\pi G_N\bar\rho\xi^2+2G_N\frac{m}{l_1}\ln\xi+\hat{B}\, , \ee
%%%%%%
where we took into account that $\triangle_{\xi}\ln \xi = 2\pi \delta\left(\vec{\xi}\, \right)$. Similar to the previous model with one noncompact dimension,
here the solution is also divergent in some directions. In the present case, it grows with the polar radius $\xi$. So, we must cut off this solution at some
distance $\xi_0$. Clearly, this boundary represents the cylindrical surface $\xi = \xi_0$. The domain in which we put the mass $m_0=m$ is a cylinder with the
radius $\xi_0$ and the generator parallel to the axis $x$. The length of the cylinder along the $x$-axis is $l_1$. The mass $m$ is in the center of the
cylinder (with the coordinate $x=0$ for the considered case). The next particle of the mass $m_1$ is inside its own cylinder with the generator along the axis
$x$ and the radius $\xi_1$. This particle may have the coordinate $x$ different from the first particle. All these cylinders have the periodic (with the period
$l_1$) boundary conditions along the $x$-axis. On the other hand, they should not overlap each other in the transverse (with respect to the $x$-axis)
direction. Moreover, it is impossible to match them smoothly via cylindrical surfaces. Therefore, we demand that the gravitational potential outside the
cylinders is equal to zero. Hence, on the boundaries of the cylinders ($\xi=\xi_0$ for the first mass) the gravitational potential and its first derivative
(with respect to $\xi$) are equal to zero. These boundary conditions enable us to determine the radius $\xi_0$ and the constant $\hat B$ in Eq. \rf{5.9}. For
example, from the condition $dC_0(\xi_0)/d\xi =0$ we get
%%%%%%
\be{5.10}
\xi_0=\sqrt{\frac{m}{\pi\bar\rho l_1}}\quad \Leftrightarrow \quad \bar\rho = \frac{m}{\pi\xi_0^2l_1}\equiv \frac{m}{s_0l_1}\, ,
\ee
%%%%%%
where $s_0=\pi\xi_0^2$ is the cross-sectional area of the cylinder. From the second boundary condition $C_0(\xi_0)=0$ we get the value of $\hat B$:
%%%%%%%
\be{5.11} \hat{B}=\frac{G_Nm}{l_1}\left(1-2\ln\xi_0\right)\, . \ee
%%%%%%%

Obviously, in the case $k_1\neq 0$, Eq. \rf{5.3} is consistent only if the following condition holds true:
%%%%%%%
\be{5.12} G_N\frac{m}{l_1}\delta\left(\vec{\xi}\, \right)+ \frac{\pi k_1^2}{l_1^2}C_{k_1}(y,z)-\frac{C''_{k_1y}(y,z)}{4\pi}-\frac{C''_{k_1z}(y,z)}{4\pi}=0\, ,
\ee
%%%%%%%
which for $\xi >0$ can be rewritten in the form
%%%%%%%
\be{5.13}
\xi\frac{d^2 C_{k_1}}{d\xi^2}+\frac{d
C_{k_1}}{d\xi}-\frac{4\pi^2 k_1^2}{l_1^2}C_{k_1}\xi=0\, .
\ee
%%%%%%%
The solutions of this equation are the modified Bessel functions:
%%%%%%
\be{5.14}
C_{k_1}(\xi)=C_1I_0\left(\frac{2\pi |k_1|}{l_1}\xi\right) - 2G_N\frac{m}{l_1} K_0\left(\frac{2\pi |k_1|}{l_1}\xi\right)\, ,
\ee
%%%%%%
where $C_1$ is a constant of integration. We took into account that the function $K_0(\xi) \to -\ln \xi$ for $\xi \to 0$, so the two-dimensional Laplacian
acting on this function provides the necessary delta-function in Eq. \rf{5.12}. The function \rf{5.14} should satisfy the same boundary conditions at
$\xi=\xi_0$ as the function $C_0(\xi)$. It can be easily verified that we cannot simultaneously reach both equalities $C_{k_1}(\xi_0)=0$ and
$dC_{k_1}(\xi_0)/d\xi =0$. Hence, we cannot determine the gravitational potential in accordance with the proposed scenario.

To conclude this section, it is worth noting that we can construct the potential in the scenario similar to the second one from the previous section. This is
the case of identical masses distributed regularly in all directions. Obviously, this case is reduced to the $T\times T\times T$ model from the section 2 with
the drawback inherent in it.

Therefore, similar to the previous section, we also failed in determining a physically reasonable gravitational potential in the model with the topology
$T\times R\times R$.

%%%%%%%%%%%%%%%%%%%%%%%%%%%%%%%%%%%%%%%%%%%%%%%%%%%%%%%%%%%%%%%%%%%%%%%%%%%%%%%%%%%%%%%%%%%%%%%%%%%%%%%%%%%%%%%%%%%%

\section{Conclusion}

\setcounter{equation}{0}

Our paper was devoted to cosmological models with different spatial topology. According to the recent observations, our Universe is spatially flat with rather
high accuracy. So, we restricted ourselves to this case. However, such spatially flat geometry may have different topology depending on a number of
directions/dimensions with toroidal discrete symmetry. These topologies result in different kinds of the lattice Universe. There are a lot of articles
exploring the lattice Universes (see, e.g., \cite{LinWh}-\cite{BruLar2} and references therein). One of their main motivations is to provide an alternative
(compared to the standard $\Lambda$CDM model) explanation of the late-time accelerated expansion of the Universe. Another important point is that our Universe
is highly inhomogeneous inside the cell of uniformity with the size of the order of 190 Mpc \cite{EZcosm2}. Hence, it is quite natural to consider the Universe
on such scales filled with discrete sources rather than a homogeneous isotropic perfect fluid.

On the other hand, $N$-body simulations of the evolution of structures in the Universe are based on dynamics of discrete sources in chosen cells. To perform
such simulations, we should know gravitational potentials generated by these sources. Therefore, the main purpose of our paper was determination of
gravitational potentials in the cases of three different spatial topologies: $T\times T\times T$, $\; T\times T\times R\; $ and $\; T\times R\times R$. The
potential satisfies the corresponding Poisson equations of the form \rf{1.1}. These equations can be obtained as a Newtonian limit of General Relativity
\cite{EZcosm2,EZcosm1}. So, to determine the potential, we should solve them. One of the main features of the analyzed Poisson equations is that they contain
the average rest-mass density which represents a constant in the comoving frame. This results in two problems. First, we cannot, in general, apply the
superposition principle. Second, the presence of such term leads to divergences in directions of noncompact dimensions. We tried to avoid these problems
arranging masses in special ways. Our investigation has shown that the $T\times T\times T$ model is the most physical one. Here, due to the discrete symmetry
in all three directions, we can represent the infinite Universe as one finite cell with periodic boundary conditions in all dimensions. The finite volume of
the cell enabled us to use the superposition principle and solve the Poisson equation for a single mass. The total potential in an arbitrary point of the cell
is equal to the sum of potentials of all particles in the cell. Unfortunately, in the case of point-like gravitating sources the obtained solution has a very
important drawback. Usually, it is expected that potentials diverge at the positions of masses. However, in the model under consideration the gravitational
potential has no definite values on the straight lines joining identical masses in neighboring cells, i.e. at points where masses are absent. Clearly, this is
a nonphysical result since the dynamics of cosmic bodies is not determined in such a case. Then, looking for a more physical solution, we smeared gravitating
masses over some regions and showed that in this case the gravitational potential has regular behavior at any point inside the cell. Therefore, smearing
represents the necessary condition of getting a regular gravitational potential in the lattice Universe. It is usually written that in $N$-body simulations
some sort of smearing is used to avoid divergence at the positions of masses. Now, we have demonstrated that this procedure helps to avoid problems on the
above mentioned straight lines as well. Therefore, the smearing of gravitating bodies in numerical simulations is not only a technical method but also a
physically substantiated procedure, and in the present paper we provide a physical justification for such smearing.

In the $T\times T\times T$ model, particles
in the cell may have different masses and be distributed arbitrarily. In the cases of topologies $\; T\times T\times R\; $ and $\; T\times R\times R$, we
cannot do this. We have shown that the only way to get a solution here is to suppose the periodic (in all dimensions) distribution of identical masses.
However, such solution is reduced to the one obtained in the case of $\; T\times T\times T\; $ topology. Therefore, first, this solution has a drawback
inherent in this model, and, second, the distribution of masses looks very artificial.

%%%%%%%%%%%%%%%%%%%%%%%%%%%%%%%%%%%%%%%%%%%%%%%%%%%%%%%%%%%%%%%%%%%%%%%%%%%%%%%%
%%%%%%%%%%%%%%%%%%%%%%%%%%%%%%%%%%%%%%%%%%%%%%%%%%%%%%%%%%%%%%%%%%%%%%%%%%%%%%%%

\section*{Acknowledgements}

The work of M. Eingorn was supported by NSF CREST award HRD-1345219 and NASA grant NNX09AV07A.

%%%%%%%%%%%%%%%%%%%%%%%%%%%%%%%%%%%%%%%%%%%%%%%%%%%%%%%%%%%%%%%%%%%%%%%%%%%%%%%%%%%%%%%
%%%%%%%%%%%%%%%%%%%%%%%%%%%%%%%%%%%%%%%%%%%%%%%%%%%%%%%%%%%%%%%%%%%%%%%%%%%%%%%%%%%%%%%

\end{document}